\documentclass{ws-procs975x65}%
\usepackage{epsfig}
\usepackage{amsmath}
\usepackage{amsfonts}
\usepackage{amssymb}
\usepackage{graphicx}%
\begin{document}
%
\title{On intrinsic invariance in Gurzadyan-Xue cosmological models}%
%

\author{H. Khachatryan, G.V. Vereshchagin and G. Yegorian}
\address{ICRANet, P.le della Repubblica 10, 65100 Pescara, Italy and \\
ICRA, Dip. Fisica, Univ. ``La Sapienza'', P.le A. Moro 5, 00185 Rome, Italy
\\E-mail: veresh@icra.it}%
%

\maketitle
%

\abstracts
{Analysis of cosmological solutions of GX dark energy models shows the presence of a separatrix
in the phase space of solutions which divides them into two classes: Friedmannian-like with initial
singularity and non-Friedmannian solutions. The invariants are found, that reveal the intrinsic symmetry
in GX models.}%

The formula for dark energy \cite{GX}
\begin{equation}
\rho_{GX}=\frac{\pi}{8}\frac{c^{4}}{G}\frac{1}{a^{2}}, \label{rhoLambda}%
\end{equation}
$c$, $G$ and $a$ are the speed of light, gravitational constant, and scale
factor of the Universe, respectively, derived by Gurzadyan and Xue predicts
the value of dark energy density in agreement with present observations
\cite{DG}. This formula leads to a possibility of variation of physical
constants such as speed of light, gravitational constant, cosmological term.
Classification of models based on the GX scaling (\ref{rhoLambda}) is given in
\cite{Ver06}. Analysis of the cosmological equations for GX models is
performed in \cite{Ver06b}.

\begin{figure}
[ptb]
\begin{center}
\includegraphics[
height=5.1407in,
width=4.846in
]%
{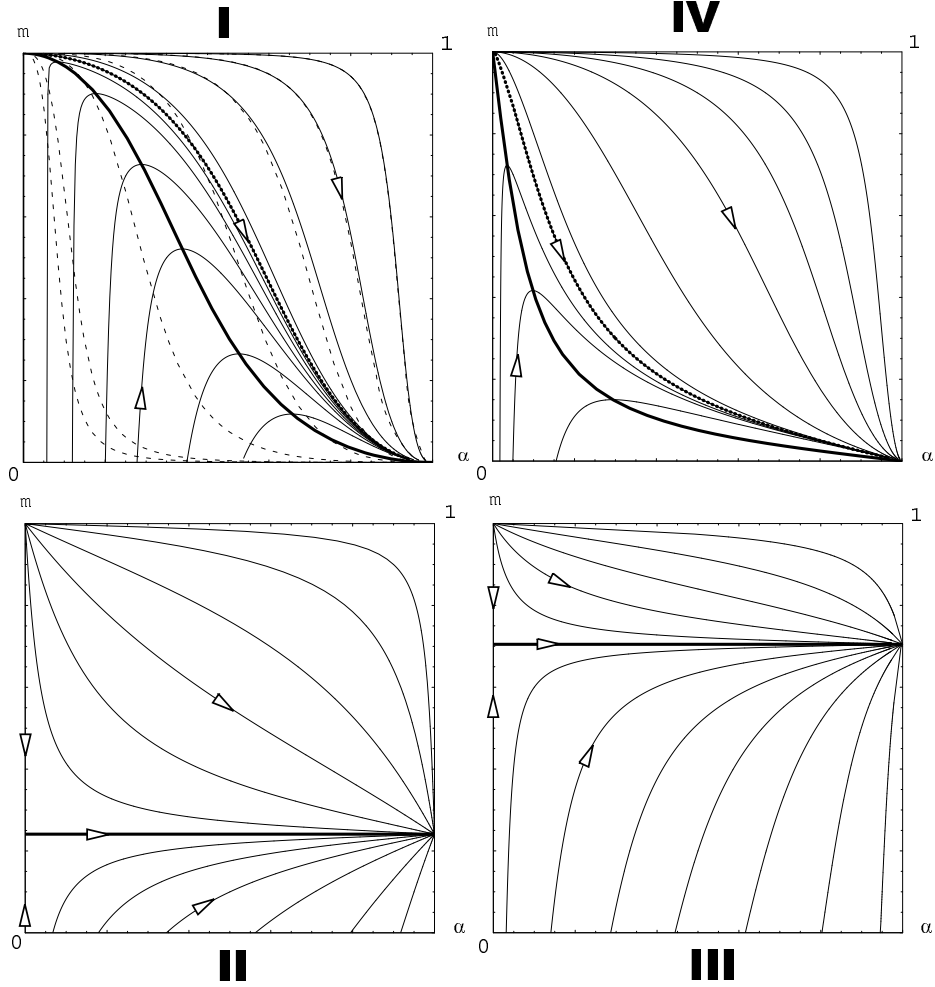}%
\caption{Phase portraits for models I-IV. Directions of phase trajectories are
shown by arrows. See detailed explanations in the text.}%
\label{pp1}%
\end{center}
\end{figure}

It was shown that each GX model is characterized by the critical value of the
density parameter $\Omega_{m}\approx2/3$ which represents a separatrix in the
space of solutions of cosmological equations \cite{Ver06a}. To illustrate the
presence of separatrix we provide in fig.\ref{pp1} phase portraits for
expansion stage with zero spatial curvature. Numeration of models is as
follows:\ I - variable dark energy density, II - variable speed of light,
constant cosmological term, III - variable gravitational constant, IV -
variable speed of light, constant dark energy density (details see e.g. in
\cite{Ver06b}).%

Phase space variables are%
\begin{equation}%
\begin{array}
[c]{l}%
\displaystyle{\alpha=\frac{2}{\pi}\text{arctg}(a)},\quad\quad\quad
\displaystyle{m=\frac{2}{\pi}\text{arctg}(\mu)},
\end{array}
\end{equation}
where $\mu$ is the matter density.

The separatrix divides the space of solutions into two
classes:\ Friedmannian-like with initial singularity and non-Friedmannian
solutions starting with vanishing density \cite{Ver06a}. The value of
separatrix is given by
\begin{equation}
\Omega_{m}=\frac{2}{3}\frac{1}{(1-k/\pi^{2})}. \label{sep}%
\end{equation}

The upper left corner in each diagram at fig.\ref{pp1} corresponds to the
Friedmann cosmological singularity with $\mu\rightarrow\infty$ and
$a\rightarrow0$. In the lower right corner $\mu\rightarrow0$ and
$a\rightarrow\infty$. The separatrix in each diagram is shown by a thick
curve. Above the separatrix solutions start with a classical singularity and
end up with infinite scale factor and zero density (or constant density for
models II and III) but the solutions below the separatrix start at a positive
scale factor and zero density and tend to the same limit.

The presence of separatrix in GX\ models is a crucial difference with the
Friedmann solutions within General Relativity. It is shown that such
symmetries are due to existence of invariant in GX models \cite{khach}; the
value of separatrix can be obtained from expression of vanishing invariant.

The continuity equation for GX models reads \cite{Ver06}
\begin{equation}
\dot{\mu}+3\frac{\dot{a}}{a}\mu=-\dot{\mu}_{GX}+(\mu+\mu_{GX})\left(
2\frac{\dot{c}}{c}-\frac{\dot{G}}{G}\right)  , \label{cont}%
\end{equation}
where $\mu_{GX}\equiv\rho_{GX}/c^{2}$ is dark energy mass density. Invariant
was found for this equation \cite{khach} which has the form%
\begin{equation}
\frac{\mu a^{3}G}{c^{2}}-\frac{{\pi}a}{4}=const\equiv b_{m}. \label{mdInv}%
\end{equation}

The value $b_{m}=0$ corresponds exactly to the above condition (\ref{sep}).

Apart from the demonstrated symmetry, present in all GX models, which is
linked with the form of (\ref{rhoLambda}), it is shown that GX models agree
well with supernovae and radio galaxies data \cite{Ver06c}. Perturbation
dynamics analysis \cite{Ver06d}\ also shows some interesting features of GX
models, in particular that density fluctuations grow also during curvature- or
vacuum-dominated stages, following matter-dominated stage of expansion, unlike
perturbations within Friedmannian models.

\end{document}